\documentclass[a4paper]{jpconf}
\usepackage{graphicx}
\usepackage{slashed}
\usepackage{amsmath}
\usepackage{amsfonts}
\usepackage{amssymb}
\usepackage{wrapfig}
\usepackage{hyperref}
\usepackage{color}
\usepackage{subfig}
\newcommand\Fontvi{\fontsize{7}{5}\selectfont}

\begin{document}

%\begin{flushright}
%%%%WITS-MITP-02x
%\end{flushright}

\title{Data driven method for predicting the shape of the dijet mass in $Z\,h + {E_{\rm T}}^{\rm {miss}}$ analysis}

\author{Skhathisomusa Mthembu$^{a,1}$, Shell May Liao$^{a,2}$, Tshidiso Molupe$^{a,3}$, Bruce Mellado Garcia$^{a,4}$, Deepak Kar$^{a,5}$}
\address{$^{a}$ School of Physics, University of the Witwatersrand, Johannesburg, Wits 2050, South Africa.}  
\ead{$^{1}$skhathisomusa.mthembu@cern.ch, $^{2}$shellmaydasilvacunha@gmail.com, $^{3}$tshidiso.sydwell.molupe@cern.ch, $^{4}$Bruce.Mellado.Garcia@cern.ch, $^{5}$Deepak.Kar@cern.ch}

\begin{abstract}
We present an analysis of new physics searches in $Z\,h$ with missing energy final states at the Large Hadron Collider considering
$Z \to l^+ l^-$ (where $l^\pm = e^\pm, \mu^\pm$) and $h \to b\bar b$ decay modes. 
For this analysis we consider production of a $CP$-odd scalar $A$ through gluon-fusion which decay into a heavy $CP$-even 
neutral scalar $H$ with $Z$-boson. 
Further $H$ decays into a lighter $CP$-even Higgs boson $h$ in association with dark matter candidate $\chi$ - a
source of missing energy. The masses of these scalars are considered as $m_h = 125$~GeV, $m_\chi = 60$~GeV, 
$2 m_h < m_H < 2 m_t$ and $m_A > 2 m_t$. A data-driven method have been applied to reduce the considerable backgrounds from
electroweak processes $W$+ jets and $Z$+ jets, in addition with top-pair and single-top production. The di-jet mass distributions have
been studied with same- and opposite-flavour lepton selections.  
\end{abstract}

\section{Introduction}
\label{intro}
In Ref.~\cite{vonBuddenbrock:2015ema}, a compatibility study with Run-I Large Hadron Collider (LHC) data at the ATLAS and the
CMS have been presented to explain the distortions in the Higgs-boson transverse momentum ($p_T$) distributions by introducing a
heavy $CP$-even scalar $H$ and a singlet neutral scalar $\chi$ in an effective theory. By proper symmetry requirement $\chi$ is
considered as a stable dark matter candidate which signatures as a source of missing transverse energy (${E_{\rm T}}^{\rm {miss}}$)
at the colliders. 
Further to these studies, in Refs.~\cite{Kumar:2016vut, vonBuddenbrock:2016psa} a proper two Higgs doublet model (2HDM)
have been introduced with neutral scalar singlets $S$ and $\chi$ where $S$ is the Higgs-like scalar. In result the enhanced particle
spectrum shall be two neutral $CP$-even scalars $h, H$ with a $CP$-odd scalar $A$, charged scalar $H^\pm$, $S$ and $\chi$ are
present in the model.
For the phenomenological implications the mass spectrum of the scalars in this model are suggested to be $m_h = 125$~GeV (as the
Standard Model (SM) Higgs), $2 m_h < m_H < 2 m_t$, $m_h \lesssim m_S \lesssim m_H - m_h$, $m_\chi < m_h/2$, 
$m_A > 2 m_t$ and $2 m_t < m_{H^\pm} < m_A$.   

In this study we take an opportunity to analyse a suggested final states with same- and opposite-flavour leptons (either $ee, \mu\mu$
or $e\mu, \mu e$) coming from $Z$-boson decay with jets (specifically with $b$-tagged jets through $h$ decay) and missing energy 
which involved the intermediate production of $A$ through gluon-fusion at the LHC and further decays 
$A \to Z H$, $H \to h \chi\chi$. In section~\ref{method}, we describe a data-driven method for new physics searches in brief following
the analysis and preliminary results in section~\ref{analysis}. We briefly summarise our work in section~\ref{summ}. 

%%%%%%%%%%%%%%%%%%%%%%%%%%%%%%%%%%%%%%%%%%%%%%%%%%%%%%%%%%%%%%%%%%%%%%%%%%%%%%%%%%%%
\section{Method Overview}
\label{method}
The ATLAS and CMS at the LHC performs a broad range of searches for new physics beyond the SM (BSM) particle spectrum which
involves a variety of final states involving jets, leptons, photons, and ${E_{\rm T}}^{\rm {miss}}$. Appropriate backgrounds will be in 
general determined with data-driven methods to perform these searches which needs to measure and control backgrounds from
QCD etc. Other backgrounds from electroweak processes like $W$+jets, $Z$+jets, and $t\bar t$ are also important as QCD 
in many BSM searches where data-driven method use to suppress such backgrounds. This method includes particular signatures of
final states like (a) selection of number of jets with minimum $p_T$ threshold, (b) determining tagging efficiencies of jets,
(c) in leptonic searches, determining the flavours with same- or opposite-signs, (d) invariant mass distributions within a mass-window,
(c) requiring large amount of ${E_{\rm T}}^{\rm {miss}}$ etc. Different observables between final states, e.g. difference of azimuthal
angle,  pseudo rapidity etc distributions also make the method highly efficient to distinct signal over the backgrounds. Also for BSM
searches a model dependent topology for a particular process within a kinematic phase space like low $p_T$ leptons and large 
number of jets are considered as inputs for data-driven method.
In this study we use this method to extract the relevant backgrounds considering the di-jet invariant mass distributions and other
criteria as discussed here.    
           
\section{Analysis and preliminary results}
\label{analysis}
For all analyses, data samples are generated from {\texttt{MadGraph}}, further decay, showering and hadronisation performed through
{\texttt{Pythia 8}} and plots, cut-flows etc are performed in {\texttt{Rivet}} and {\texttt{ROOT}} framework. 

The process and parameter choices for our studies are as follows:
\begin{itemize}
\item $p p \to g g \to A \to Z H$ at $\sqrt{s} = 13$~TeV, with $m_A = 750$~GeV, $m_H \in [260, 350]$~GeV,
\item $Z \to l^+ l^-$, where $l^\pm = e^\pm, \mu^\pm$,
\item $H \to h \chi\chi$, with $m_\chi = [50, 60]$ GeV,
\item $h \to b\bar b$, with 70\% $b$-tagging efficiencies.  
\end{itemize}
We started our analysis by optimising the signal sample over relevant sources of backgrounds from $W$+ jets, $Z$+ jets, di-bosons,
top-pair and single-top samples, where the cross sections of these backgrounds are normalised with available yield samples at the
ATLAS frameworks. Here all jets are reconstructed with anti-$k_T$ algorithm within $\Delta R = 0.4$, where 
$\Delta R = \sqrt{ \left( \Delta\phi \right)^2 + \left(\Delta\eta \right)^2 }$ is the jet separation radius with $\phi$ and $\eta$ as the
azimuthal angle and pseudo rapidity of jets. The three major selection criteria for optimisation are:
\begin{itemize}
\item[(1)] Selection of same- and opposite-flavour leptons,
\item[(2)] At least two central jets with $b$-tagging, and
\item[(3)] Identify exactly one- and at least two- $b$-tagged jets. In case of at least two $b$-tagged jets, one must have $p_T > 45$~GeV. 
\end{itemize}     
Other conditions and requirements are:
\begin{itemize}
\item[(a)] Leptons are required with $p_T > 25$~GeV,
\item[(b)] Invariant mass of leptons, $m_Z (m_{ll})\in [65, 115]$~GeV,
\item[(c)] $E_{\rm T}^{\rm miss} > 100$~GeV and
\item[(d)] The azimuthal angle separation within leading (sub-leading) jets in $p_T$-ordering with $E_{\rm T}^{\rm miss}$ should be
$\Delta\phi_{j_{1(2)}, E_{\rm T}^{\rm miss}} > 0.5$.   
\end{itemize}     
After applying these selections over the samples, we tabulated the cut flows in terms of weighted events for each cut at $\sqrt{s} = 13$~TeV with integrated luminosity of 3.32 ${\rm fb}^{-1}$ in Table~\ref{tab1} and~\ref{tab2}.  
\begin{center}
\begin{table}[h]
	\caption{\label{tab1}Number of weighted events per cut, over different samples, for the same-flavour scheme.}
	\Fontvi
	\begin{tabular}{ l | l l l l l }
		\hline \hline 
		$ee/\mu \mu$ & SingleTop & Diboson & W+Jets & Z+Jets & t$\bar{t}$ \\ \hline \hline 
				
		No Cuts & 9790.13  $\pm$ 16.85 & 6579.72  $\pm$ 75.22 & 43970.93  $\pm$ 660.50 & 1883642.25  $\pm$ 4824.70 & 107052.43  $\pm$ 99.49  \\ 
		Opposite Charge & 8588.12  $\pm$ 15.70 & 6366.03  $\pm$ 74.13 & 24653.45  $\pm$ 478.14 & 1861689.12  $\pm$ 4804.25 & 99578.14  $\pm$ 95.95  \\ 
		Same Flavour & 4309.85  $\pm$ 11.12 & 6166.61  $\pm$ 72.80 & 11658.15  $\pm$ 323.77 & 1843588.00  $\pm$ 4791.81 & 49938.71  $\pm$ 67.96  \\
		Lepton $p_{T}$ & 2413.90  $\pm$ 8.27 & 4344.61  $\pm$ 61.08 & 890.54  $\pm$ 84.10 & 1347672.12  $\pm$ 4340.26 & 29059.95  $\pm$ 51.84  \\
		$m_{ll}$ & 639.91  $\pm$ 4.26 & 4191.63  $\pm$ 60.12 & 182.57  $\pm$ 45.40 & 1281533.62  $\pm$ 4287.27 & 8188.96  $\pm$ 27.53  \\
		At least 2 jets & 435.30  $\pm$ 3.51 & 3422.53  $\pm$ 54.76 & 130.24  $\pm$ 32.40 & 500706.16  $\pm$ 2484.59 & 7443.30  $\pm$ 26.24 \\ \hline
 		${E_T}^{miss} > 100$ GeV & 89.15  $\pm$ 1.59 & 11.19  $\pm$ 2.39 & 12.57  $\pm$ 4.17 & 706.34  $\pm$ 55.94 & 1571.01  $\pm$ 12.04 \\ \hline
		$\Delta \phi (j1, {E_T}^{miss}) > 0.5$ & 87.76  $\pm$ 1.58 & 9.28  $\pm$ 1.76 & 12.45  $\pm$ 4.16 & 570.42  $\pm$ 52.97 & 1541.29  $\pm$ 11.93  \\ 
		$\Delta \phi (j2, {E_T}^{miss}) > 0.5$ & 79.67  $\pm$ 1.50 & 3.40  $\pm$ 0.98 & 12.14  $\pm$ 4.16 & 447.34  $\pm$ 49.35 & 1368.99  $\pm$ 11.24 \\ \hline 
		Exactly 1b & 42.97  $\pm$ 1.10 & 1.06  $\pm$ 0.62 & 0.46  $\pm$ 0.27 & 24.93  $\pm$ 4.20 & 649.61  $\pm$ 7.75  \\ 
		Exactly 2b & 16.11  $\pm$ 0.68 & 0.00  $\pm$ 0.00 & 0.04  $\pm$ 0.02 & 2.37  $\pm$ 1.18 & 490.09  $\pm$ 6.72  \\ \hline 
		b-jet $p_{T} > 45 GeV$ & 15.04 $\pm$ 0.65 & 0.00  $\pm$ 0.00 & 0.03  $\pm$ 0.02 & 1.56  $\pm$ 0.96 & 462.16  $\pm$ 6.53  \\
		\hline \hline 
		
	\end{tabular}
\end{table}
\end{center}	
	
\begin{figure}[!htbp]
	\centering
	\subfloat[]{\includegraphics[width=0.5\textwidth,height=0.4\textwidth]{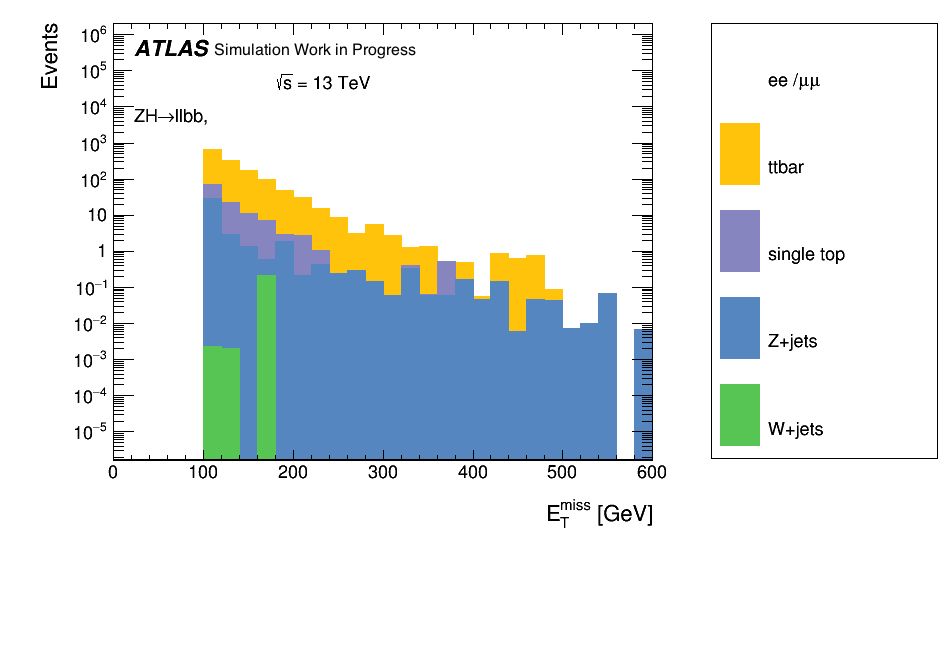}} 
	\subfloat[]{\includegraphics[width=0.5\textwidth,height=0.4\textwidth]{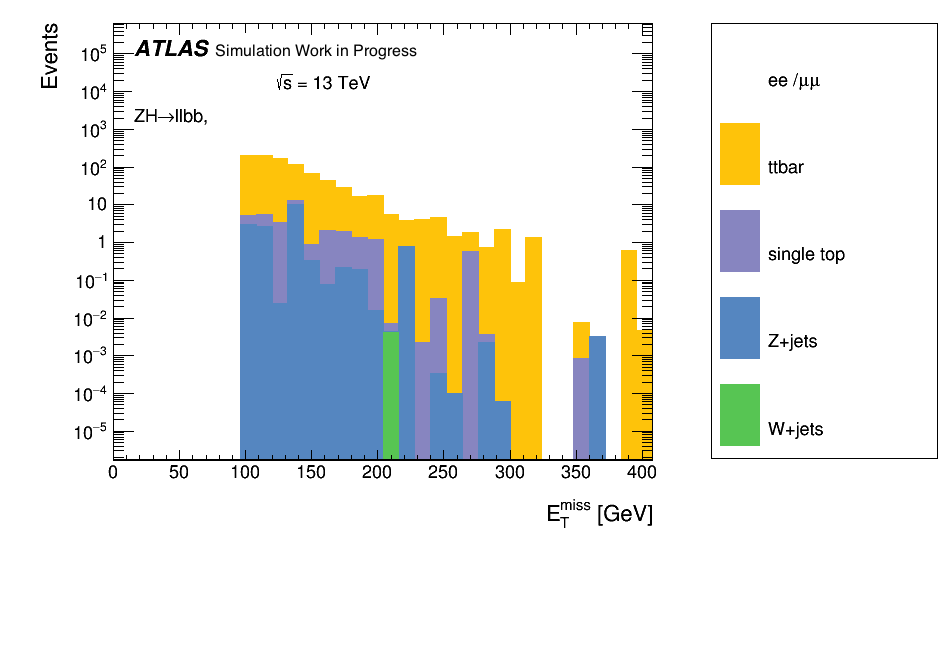}} \\
	\subfloat[]{\includegraphics[width=0.5\textwidth,height=0.4\textwidth]{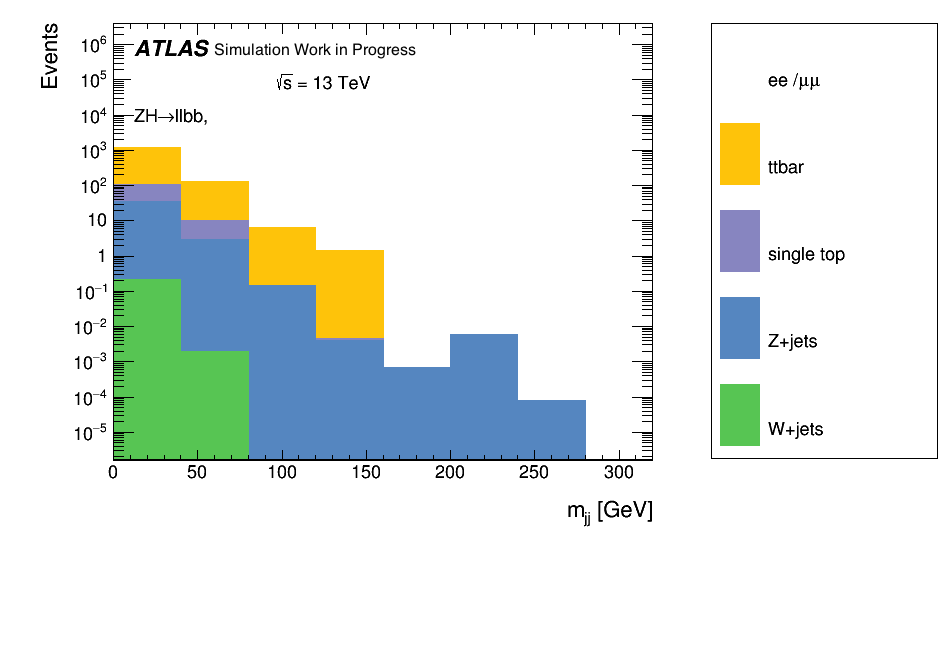}} 
	\subfloat[]{\includegraphics[width=0.5\textwidth,height=0.4\textwidth]{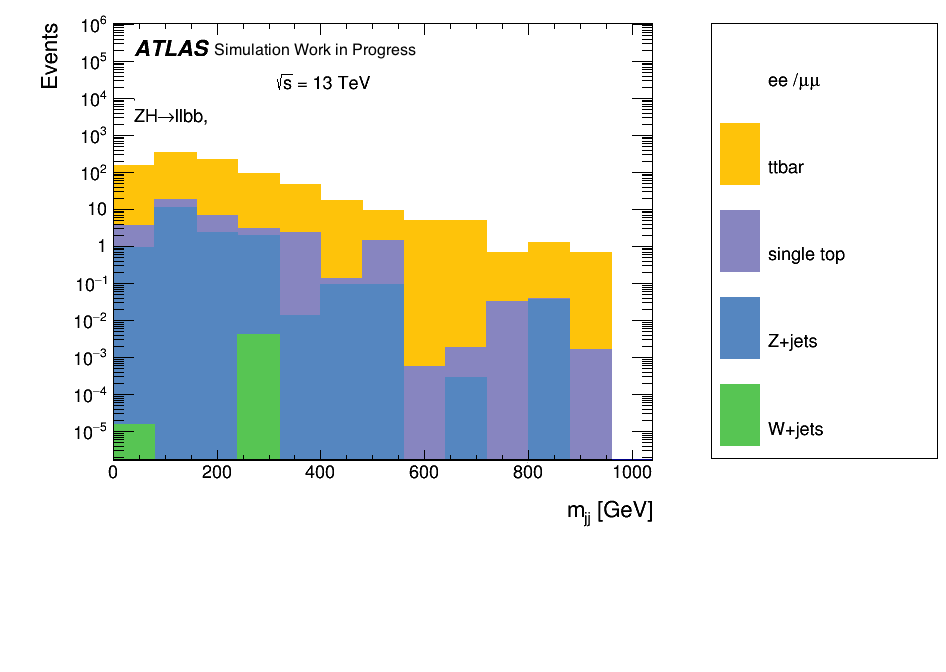}} \\
	\subfloat[]{\includegraphics[width=0.5\textwidth,height=0.4\textwidth]{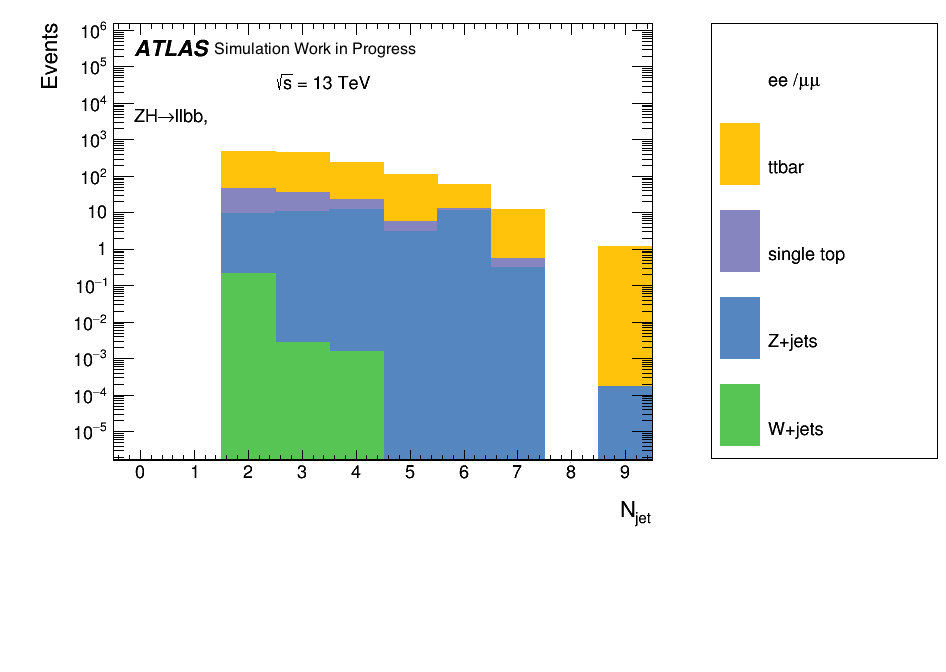}} 
         \subfloat[]{\includegraphics[width=0.5\textwidth,height=0.4\textwidth]{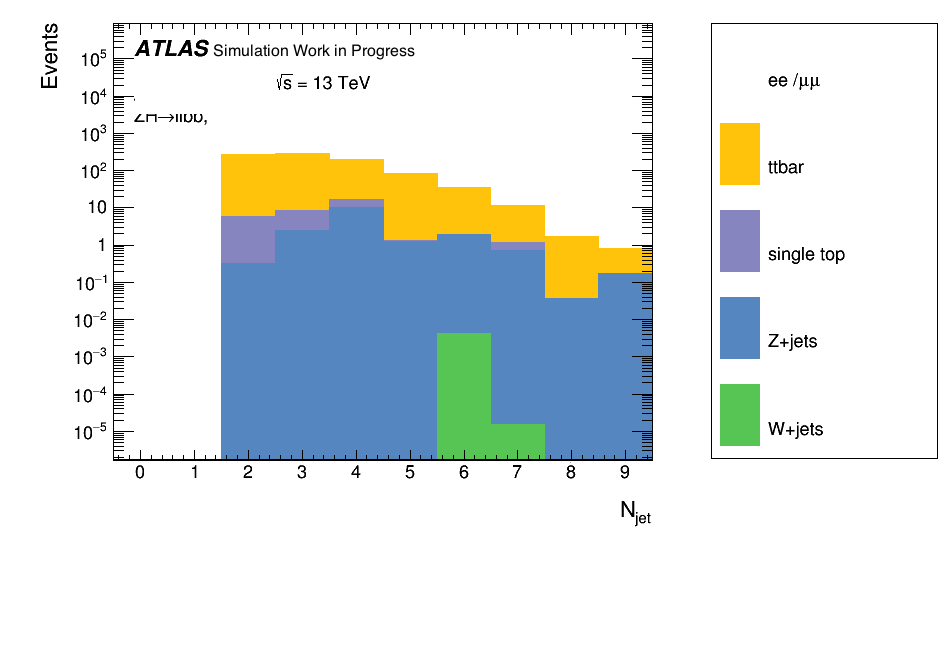}} 
	\caption{Normalised distributions of (a), (b) missing transverse momentum, (c), (d) di-jet mass spectrum and (e), (f) jet multiplicities 
	for exactly one, at least two $b$-tag selections under the same-flavour scheme.}
	\label{fig:a}
\end{figure}

\begin{figure}[!htbp]
	\centering
	\subfloat[]{\includegraphics[width=0.5\textwidth,height=0.4\textwidth]{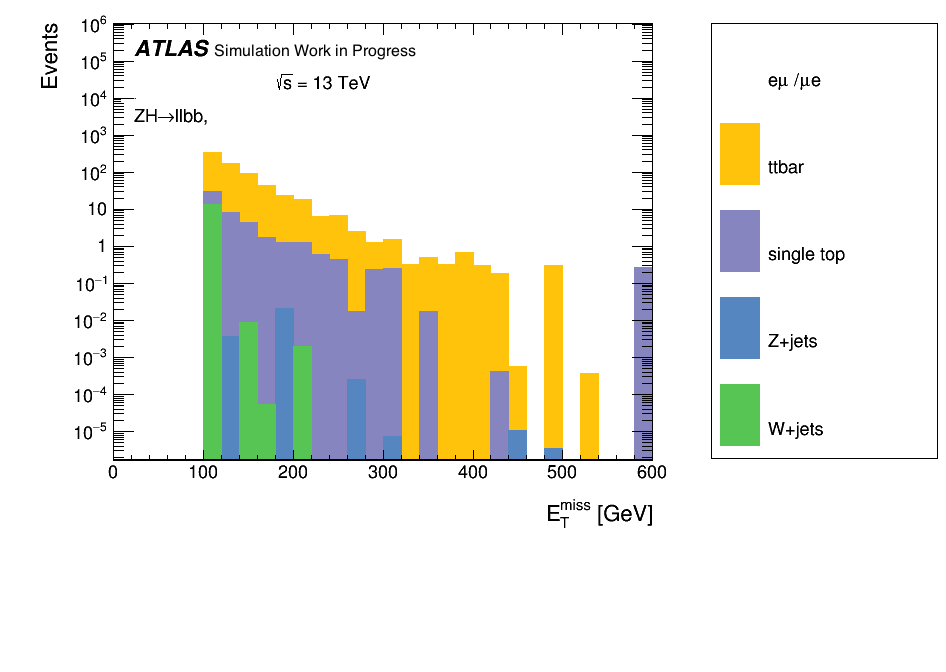}} 
	\subfloat[]{\includegraphics[width=0.5\textwidth,height=0.4\textwidth]{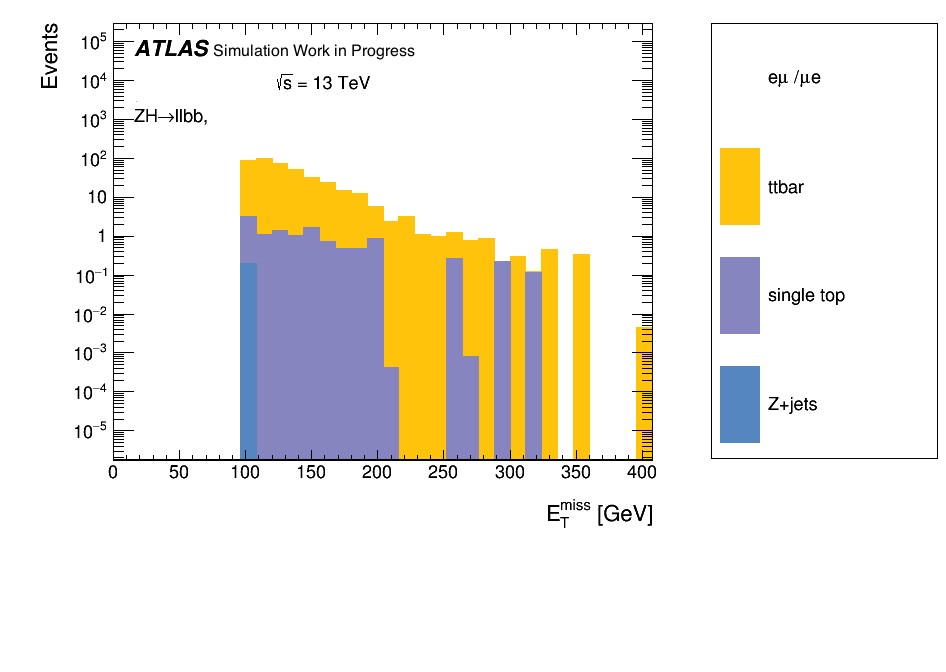}} \\
	\subfloat[]{\includegraphics[width=0.5\textwidth,height=0.4\textwidth]{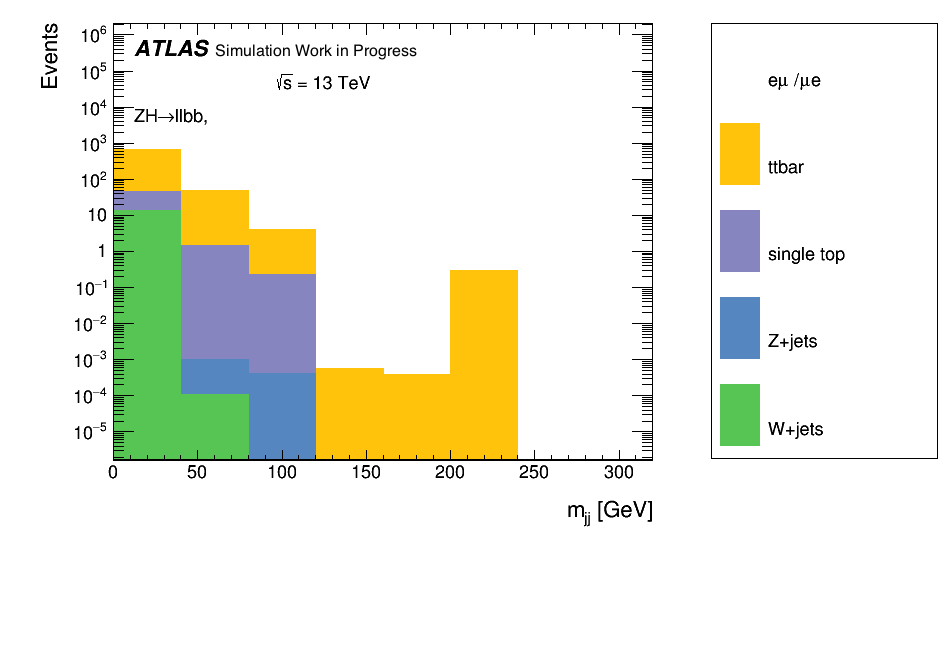}} 
	\subfloat[]{\includegraphics[width=0.5\textwidth,height=0.4\textwidth]{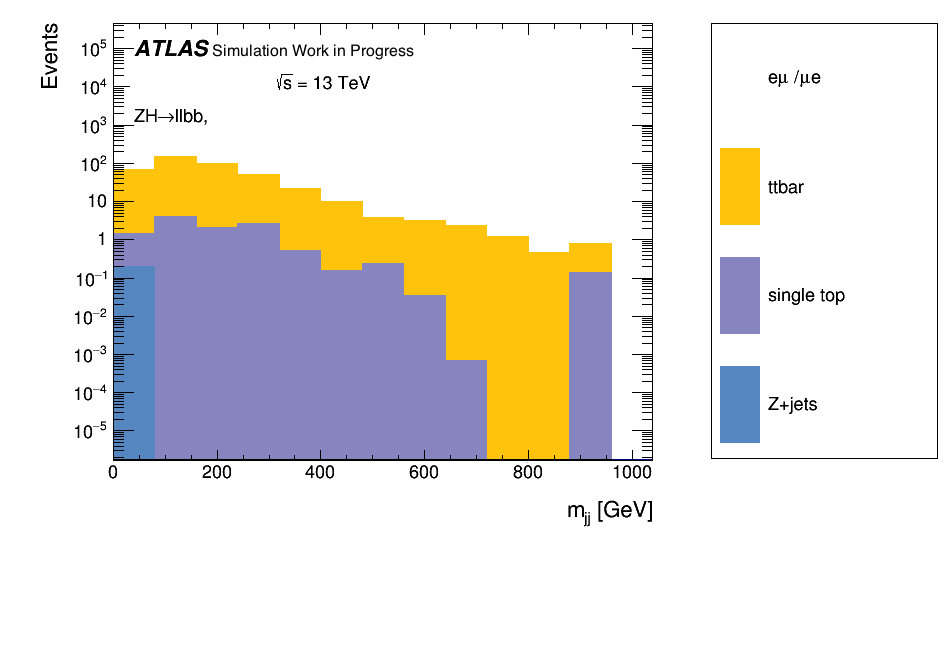}} \\
	\subfloat[]{\includegraphics[width=0.5\textwidth,height=0.4\textwidth]{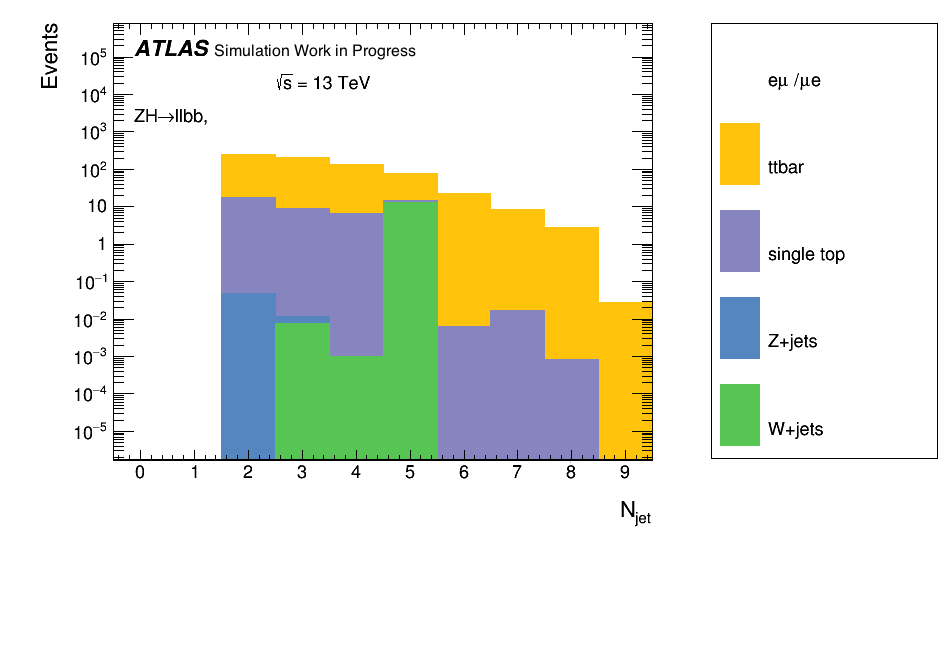}} 
	\subfloat[]{\includegraphics[width=0.5\textwidth,height=0.4\textwidth]{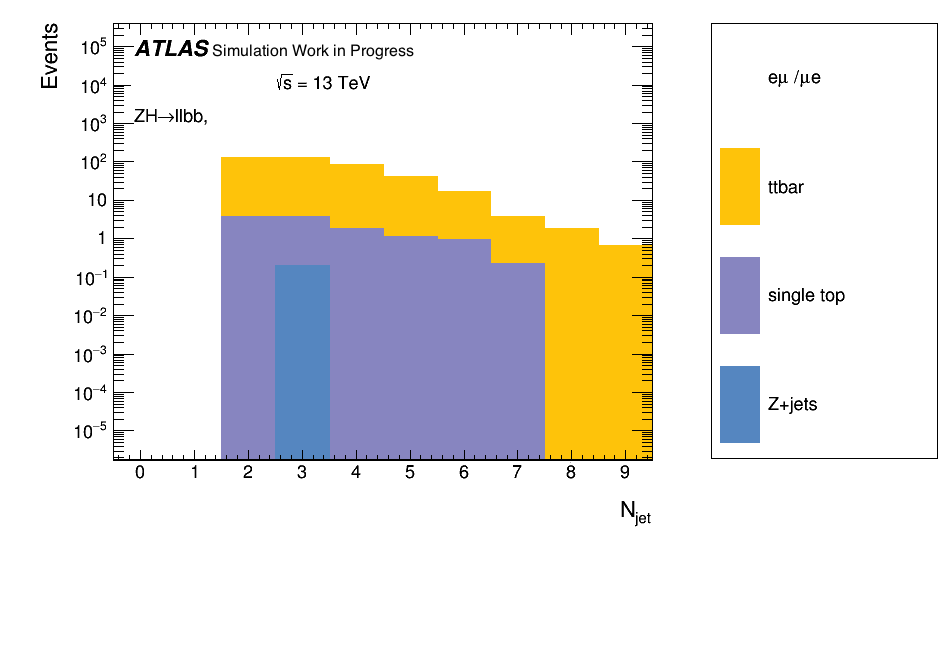}} 
	\caption{Normalised distributions of (a), (b) missing transverse momentum, (c), (d) di-jet mass spectrum and (e), (f) jet multiplicities 
	for exactly one, at least two $b$-tag selections under the opposite-flavour scheme.}
	\label{fig:b}
\end{figure}

\begin{center}
\begin{table}[h]
			\caption{\label{tab2}Number of weighted events per cut, over different samples, for the different-flavour scheme.}
			\Fontvi
		\begin{tabular}{ l | l l l l l }
			\hline \hline 
			 \textbf{$e \mu /\mu e$}  & SingleTop & Diboson & W+Jets & Z+Jets & $t \bar{t}$\\
			
			\hline \hline
			
			No Cuts& 9790.13  $\pm$ 16.85 & 6579.72  $\pm$ 75.22 & 43970.93  $\pm$ 660.50 & 1883642.25  $\pm$ 4824.70 & 107052.43  $\pm$ 99.49  \\ 
			Opposite Charge & 8588.12  $\pm$ 15.70 & 6366.03  $\pm$ 74.13 & 24653.45  $\pm$ 478.14 & 1861689.12  $\pm$ 4804.25 & 99578.14  $\pm$ 95.95  \\ 
			Different Flavour & 4278.26  $\pm$ 11.08 & 199.42  $\pm$ 13.98 & 12995.30  $\pm$ 351.84 & 18101.14  $\pm$ 345.59 & 49639.43  $\pm$ 67.74  \\ 
			Lepton $p_{T} > 25 GeV$ & 2387.51  $\pm$ 8.22 & 41.22  $\pm$ 6.79 & 1086.84  $\pm$ 107.67 & 7299.90  $\pm$ 225.93 & 28852.22  $\pm$ 51.65 \\ 
			$m_{ll}$ & 638.78  $\pm$ 4.26 & 13.41  $\pm$ 3.57 & 247.19  $\pm$ 56.10 & 5688.96  $\pm$ 206.32 & 8119.21  $\pm$ 27.40  \\ 
			At Least 2 jets & 438.17  $\pm$ 3.53 & 8.56  $\pm$ 3.10 & 189.05  $\pm$ 46.39 & 2288.16  $\pm$ 124.05 & 7368.57  $\pm$ 26.10  \\ \hline 
			${E_{T}^{miss}}$ & 88.70  $\pm$ 1.59 & 1.40  $\pm$ 0.74 & 8.03  $\pm$ 4.06 & 38.19  $\pm$ 9.94 & 1569.08  $\pm$ 12.04  \\ \hline 
			$\Delta \phi (j1, {E_{T}^{miss}}) > 0.5$ & 87.11  $\pm$ 1.57 & 0.77  $\pm$ 0.47 & 7.99  $\pm$ 4.06 & 16.28  $\pm$ 8.26 & 1542.07  $\pm$ 11.93 \\ 
			$\Delta \phi (j2, {E_{T}^{miss}}) > 0.5$ & 78.80  $\pm$ 1.49 & 0.36  $\pm$ 0.24 & 7.52  $\pm$ 4.05 & 7.40  $\pm$ 6.29 & 1364.80  $\pm$ 11.22  \\ \hline 
			Exaclty 1 b jet & 43.44  $\pm$ 1.11 & 0.19  $\pm$ 0.18 & 0.67  $\pm$ 0.27 & 1.29  $\pm$ 0.57 & 657.72  $\pm$ 7.79  \\ 
			At least 2 b jets & 16.23  $\pm$ 0.68 & 0.00  $\pm$ 0.00 & 0.02  $\pm$ 0.02 & 0.06  $\pm$ 0.11 & 480.07  $\pm$ 6.65  \\ \hline 
			b-jet $p_{T} > 45 GeV$ & 15.23 $\pm$ 0.66 & 0.00  $\pm$ 0.00 & 0.02  $\pm$ 0.02 & 0.00  $\pm$ 0.00 & 449.80 $\pm$ 6.44  \\
			\hline	\hline 
			
		\end{tabular}
\end{table}		
\end{center}	
In Figures~\ref{fig:a}(~\ref{fig:b}), we show normalised distributions of missing transverse momentum, invariant di-jet mass spectrum and
jet multiplicities for two selection criteria (1) exactly one $b$-tagged jets and (2) at least two $b$-tagged jets under the same-(opposite-) flavour schemses. 

\begin{figure}[!htbp]
	\centering
%	\subfloat[]{\includegraphics[width=0.5\textwidth,height=0.4\textwidth]{figures/C1_0ptag0pjet_0ptv_SR_exactly1b_MET_Log}} 
%	\subfloat[]{\includegraphics[width=0.5\textwidth,height=0.4\textwidth]{figures/C1_0ptag0pjet_0ptv_SR_exactly2b_MET_Log}} \\
	\subfloat[]{\includegraphics[width=0.5\textwidth,height=0.4\textwidth]{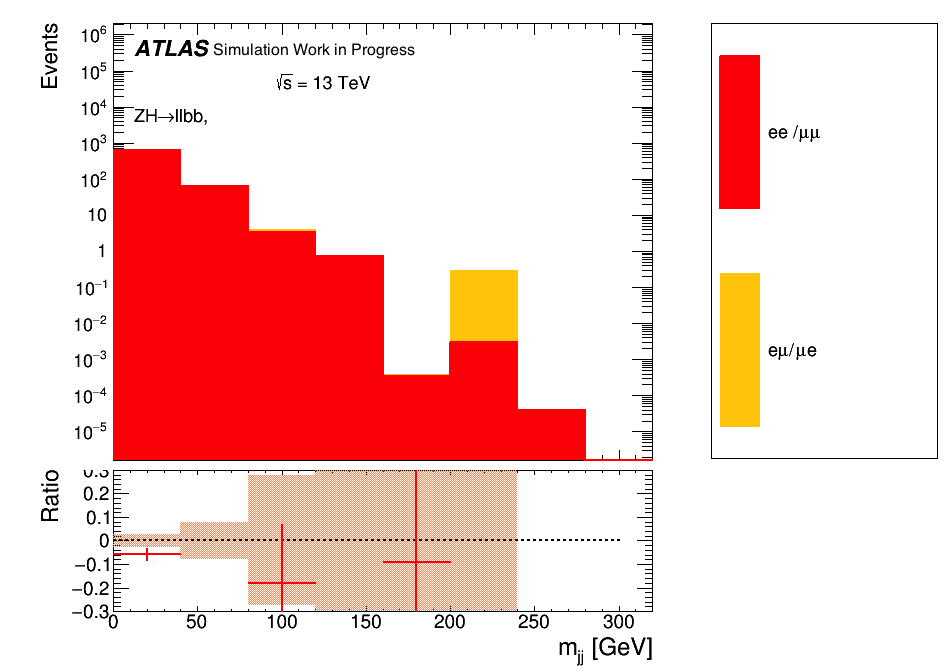}} 
	\subfloat[]{\includegraphics[width=0.5\textwidth,height=0.4\textwidth]{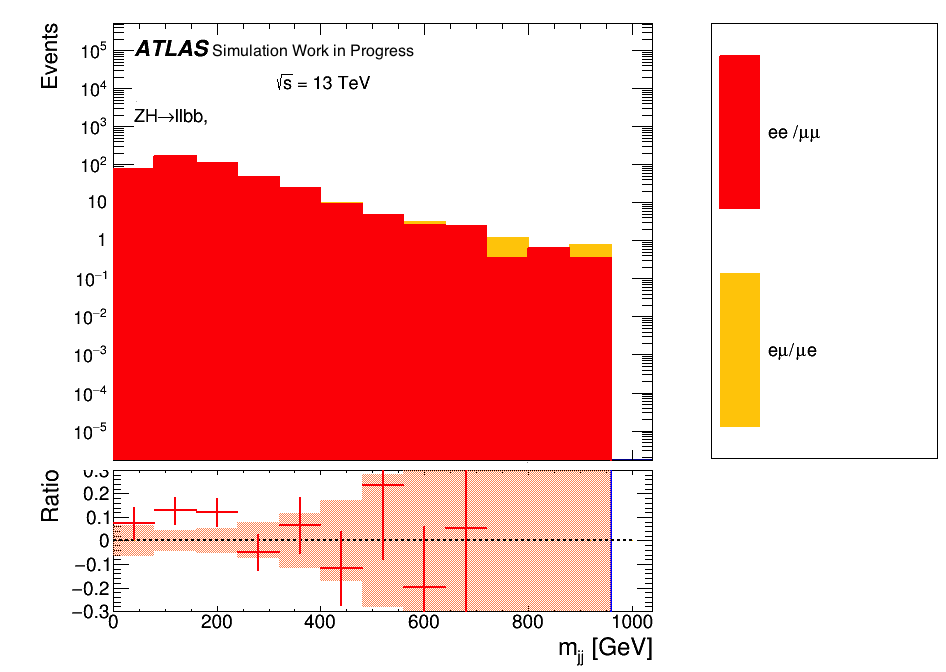}} 
	\caption{Normalised ratio distributions of di-jet mass spectrum  
	for (a) exactly one and (b) at least two $b$-tag selections under the same- and opposite-flavour scheme.}
	\label{fig:c}
\end{figure}

We draw a ratio distribution of invariant di-jet mass for exactly one and at least two $b$-tagged selection criteria for the same and 
opposite flavour leptons in Figure~\ref{fig:c} and in result we see the difference between two schemes are negligible, which shows our data driven method
suffice for the persuaded searches. 

\section{Summary}
\label{summ}
As stated, the objective is to compare flavour schemes for $Z$ boson decay leptons, and employing a data driven method to predict the spectrum of the di-jet mass. To do this, cuts were enforced onto the weighted sample event numbers to suppress the background. After significant cuts, the number of background events were lowered substantially, to a point where the events for the same flavour and different flavour schemes were not only comparable - as seen in the cut flows, but also relatively close - as seen in the ratio plots shown. The suppression of the $Z$ + jets events is observed on the distributions shown.  Very few statistics were observed after the 2 $b$-tag cut, but the relative difference between the flavour schemes is notably lowered after at the $b$ cuts.

\section*{Acknowledgements}
We acknowledge High Energy Physics Group at the University of the Witwatersrand for fruitful discussions on the subject. SM, SML and 
TM acknowledges funding support from National Research Foundation, South Africa for financial support towards the research work.

\end{document}